KAPSARC
مركز الملك عبدالله للدراسات والبحوث البترولية
King Abdullah Petroleum Studies and Research Center

Discussion Paper

# Driving Reductions in Emissions

## Unlocking the Potential of Fuel Economy Targets in Saudi Arabia

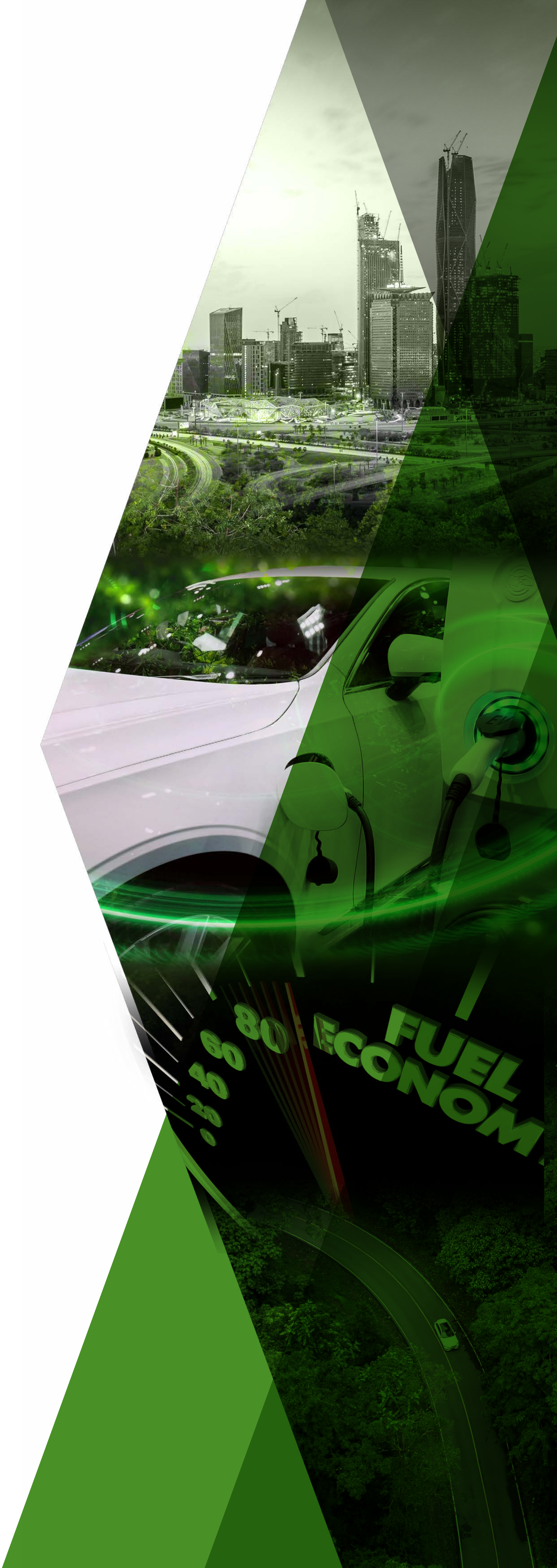

Ibrahem Shatnawi and Jeyhun I. Mikayilov

## About KAPSARC

KAPSARC is an advisory think tank within global energy economics and sustainability providing advisory services to entities and authorities in the Saudi energy sector to advance Saudi Arabia's energy sector and inform global policies through evidence-based advice and applied research.

This publication is also available in Arabic.

## Legal Notice

# Abstract

The adoption of more stringent fuel economy standards represents a pivotal pathway toward achieving net zero emissions in the transportation sector. By steadily increasing the fuel efficiency of vehicles, this approach drives a gradual but consistent decline in emissions. When coupled with the simultaneous integration of electric and alternative fuel vehicles into the market, the goal of net zero emissions becomes increasingly feasible.

In this study, we demonstrated the impact of improving fuel economy standards, mainly for light-duty vehicles (gasoline-based), within Saudi Arabia's transportation sector. The analysis, rooted in the use of standard logistic functions to make projections for car sales and the AFLEET tool, elucidates the transformative impact of adhering to the proposed fuel economy standards (CAFE standard). Should Saudi Arabia implement the phase-in compliance method aligned with the policy update (ICCT 2014) regarding the corporate average fuel economy standard for new light-duty vehicles, the actual fleet average (CAFE) would specifically pertain to the proportion of sales associated with its most efficient vehicles. Thus, the analysis in this paper is specifically focused on light-duty vehicles and is limited to annual car sales data.

According to the estimations for 2016-2020, adhering to the CAFE standards proposed in 2016 could have decreased both the cumulative energy consumption and greenhouse gas (GHG) emissions for this period by 20%. Furthermore, to address fuel economy improvement targets, the cumulative energy consumption and GHG emissions for 2021 to 2030 could decrease by an additional 10% under scenario 1. Scenario 3 (assuming 3% efficiency per annum from 2021 to 2030), in contrast to scenario 1, shows a marginal decrease of approximately 1.2% in both energy consumption and GHG emissions reduction. Additionally, from 2016 to 2020, this scenario could have decreased overall GHG emissions from the transportation sector by approximately 1.6%. Furthermore, by 2030, GHG emissions could decline by an additional 4.16% if scenario 3 is enacted.

# 1. Introduction

In an era of increased concern about climate change and the sustainability of our planet's resources, the role of the automobile sector in defining the trajectory of energy use and environmental effects is critical. In 2022, global transportation $CO_2$ emissions increased by 3% compared to those in the previous year (IEA 2023). From 1990 to 2022, transportation emissions worldwide increased at an annual average pace of 1.7%, which was faster than that of any other end-use sector (IEA 2023). Solid policies, fiscal incentives, and significant investment in infrastructure for sustainable vehicle operations will be necessary to meet worldwide emissions reduction targets. As the worldwide demand for transportation grows, the need for more fuel-efficient cars becomes more important.

The introduction of fuel economy goals – a set of regulatory requirements aiming to improve vehicle efficiency and lower the carbon footprint – is one of the primary policy instruments used to address this dilemma.

Fuel economy policies such as the Corporate Average Fuel Economy (CAFE), European Union $CO_2$ Emission Standards, and China's New Energy Vehicle Credit System have gained significance worldwide as countries and international organizations work to reduce the negative consequences of energy usage and greenhouse gas (GHG) emissions. These policies dictate certain standards for fuel per unit of miles traveled that vehicle manufacturers must meet and thus prevent the sale of vehicles that fail to meet this standard. Such regulations are intended to minimize the economic burden of growing fuel prices and the environmental outcomes of fossil fuel consumption.

In 2019, the road transportation sector in the Kingdom of Saudi Arabia (KSA), similar to many countries globally, contributed substantially to GHG emissions; it represented 21% of the total carbon dioxide ($CO_2$) emissions (Akinpelu 2023). In the KSA, gasoline-based engines are still the predominant type used to power light-duty vehicles. An internal combustion engine (ICE) has historically been the preferred choice for these vehicles; this preference is expected to continue until at least 2040 (ARAMCO 2023).

With respect to reducing GHG emissions through the efficient use of energy, several initiatives have been implemented in the KSA; for instance, different energy price management programs have been in place since 2014. Numerically, the price of 91-octane gasoline increased by 481% from 2014 (0.375 SAR/L) to 2023 October (2.18 SAR/L). The corresponding increase in the 91-octane gasoline price was 288% from 2014 (0.60 SAR/L) to October 2023 (2.33 SAR/L). The KSA, which is relatively new to implementing fuel economy standards, has introduced yearly fee-based incentives tied to vehicle fuel efficiency, signaling a commitment to sustainability (SPA 2023).

Additionally, to increase the efficiency of energy use in transport, several social awareness programs have been implemented to educate automobile owners. Moreover, the KSA has launched fuel economy standards and updated them several times (Shannak et al. 2024).

This study investigated the interrelationship between fuel efficiency targets, energy conservation, and environmental well-being in the Saudi context. The main goal was to examine the impact of these targets on energy use and emissions. The structure of this paper includes a background section that thoroughly summarizes global fuel economy targets and their international evolution, as well as the progress made in the KSA. The model section is used to examine the environmental consequences of different fuel economy targets, and it presents the empirical findings. Ultimately, the paper concludes by summarizing the findings and presenting policy implications derived from the insights gained via the study.

# 2. Background

The transportation sector in the KSA represents a dynamic and evolving landscape that plays a pivotal role in the country's economic development and connectivity. The KSA has made significant progress in achieving sustainable and diversified economic development in the transport sector through its Vision 2030 projects. Recent years have witnessed substantial investments in expanding and enhancing public transportation, exemplified by the introduction of metro systems in Riyadh city as well as the expansion of the public transport bus network in major cities such as Makkah, Jeddah, Madinah, and Dammam. These programs have fostered sustainable and efficient mobility options. The aviation sector remains robust, boasting modern international airports and a growing fleet of airlines to facilitate domestic and international travel. Maritime transport is vital for the import and export of goods and is facilitated by strategically located seaports (Oxford Business Group 2022). As part of the Vision 2030 initiative, the KSA is actively promoting sustainable transportation solutions, with a focus on reducing environmental impact and fostering innovation in the sector.

The KSA has taken a significant step toward environmental sustainability and combating climate change by demonstrating its unwavering commitment to the Paris Agreement (UNFCCC 2021). Under this globally recognized accord, nations worldwide have united in their efforts to mitigate the effects of climate change by reducing GHG emissions and fostering a cleaner, more sustainable future.

Furthermore, the KSA has demonstrated its dedication to preserving the environment. On November 17, 2014, prior to the KSA committing to the Paris Agreement, the Saudi Standards, Metrology and Quality Organization (SASO) released new fuel economy rules for light-duty vehicles (LDVs) in the KSA (ICCT 2014). These guidelines were intended to reduce the amount of GHG emissions produced by LDVs. The proposed regulations apply to all passenger vehicles and light trucks.

Figure 1 depicts a comparative analysis between the standards of the KSA and the standards that have been implemented or suggested by various countries globally (ICCT 2023).

As shown in Figure 1, each country is navigating its unique path to promote fuel efficiency and reduce GHG emissions from light-duty vehicles. While historical trends reflect gradual progress, future projections hint at more aggressive targets, increased adoption of electric and hybrid vehicles, and stringent regulations to align with global sustainability goals. The transition toward cleaner and more efficient transportation is a common thread across these diverse geopolitical contexts, emphasizing a collective commitment to a greener automotive future.

In the KSA, as shown in Figure 1, the trajectory toward enhanced fuel efficiency and reduced GHG emissions for light-duty vehicles follows a common theme of gradual progress and evolving standards compared with other countries such as the United States, Canada, and the U.K. For example, in the United States and Canada, historical trends have seen a steady increase in fuel efficiency standards and a parallel push to lower GHG emissions from vehicles. Future projections indicate a continued focus on these goals, potentially with more ambitious

**Figure 1.** Passenger car emissions and consumption, normalized to the new European driving cycle (NEDC).

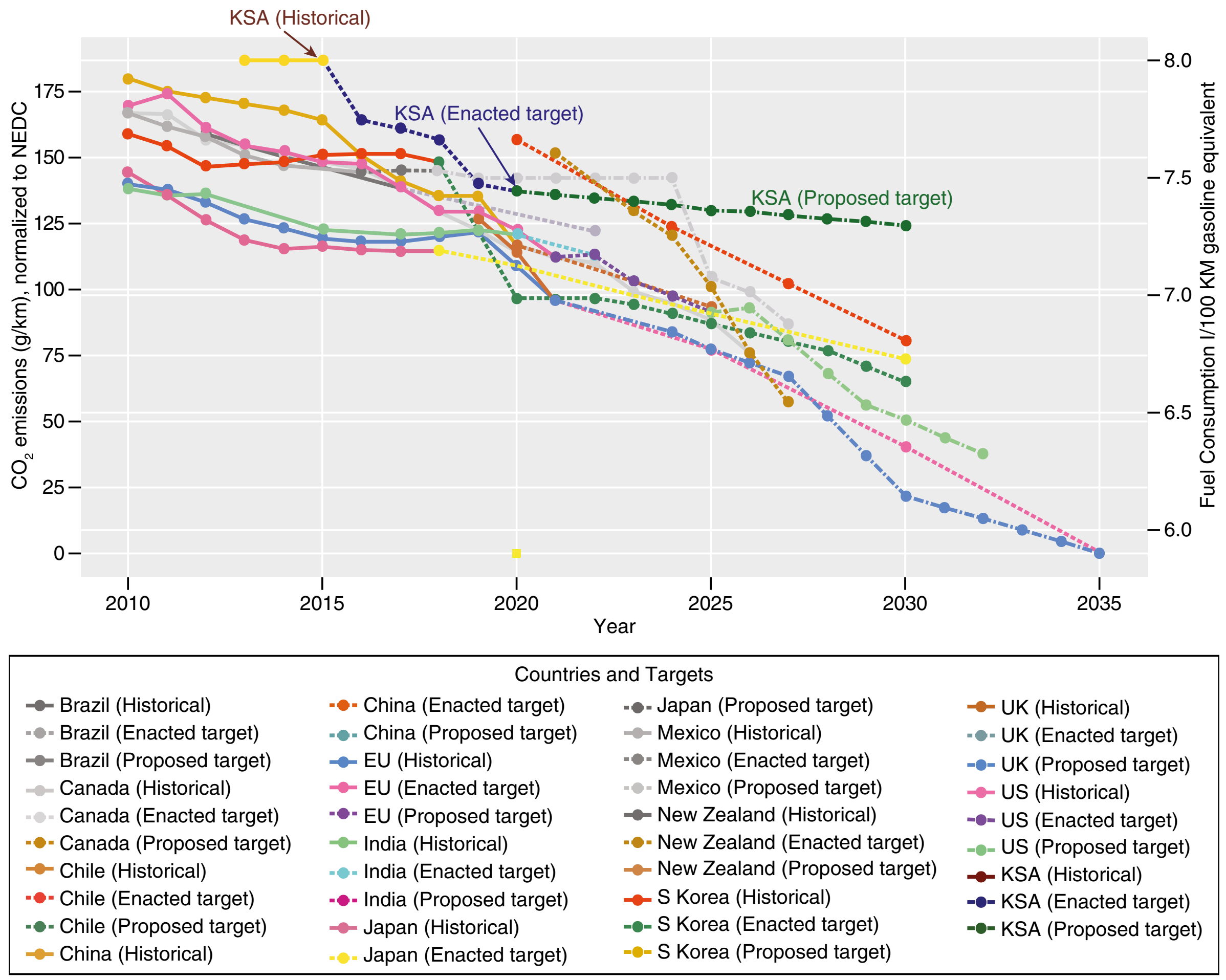

*Source: ICCT (2023) and Authors' Interpretation.*

targets aligned with national climate objectives. The U.K., known for its aggressive environmental policies, has progressively implemented stringent standards and intends to ban the sale of new fossil fuel vehicles by 2035 while supporting electric vehicle adoption (Roberts 2023). Future projections suggest an accelerated move toward electric vehicle dominance, which aligns with ambitious emission reduction targets and decreased reliance on traditional combustion engines.

Saudi Arabia has introduced a new policy in which vehicle fees are based on fuel efficiency standards (SPA 2023); the intent is to reduce carbon emissions and increase energy efficiency in road transport. This policy imposes annual vehicle fees that are determined by fuel economy standards. The objective of this policy is to promote the adoption of vehicles that are more fuel-efficient. This step is in accordance with Saudi Arabia's dedication to the Paris Agreement. By incentivizing fuel efficiency through a fee structure, Saudi Arabia is encouraging both manufacturers and consumers to embrace eco-friendly technologies and adopt vehicles that are kinder to the environment. This initiative highlights Saudi Arabia's determination to transition toward a low-carbon economy, setting the stage for a more sustainable, environmentally conscious future.

The proposed economic instrument is an annual fee added when issuing and renewing a vehicle's license; its value ranges between 0 and 190 SAR (SPA 2023). The aim of this fee is to stimulate the trend toward owning more fuel-efficient vehicles and thus reduce harmful emissions and preserve the environment and natural resources of the KSA. Economical vehicles with high fuel efficiency are exempt from this consideration. In its decision, the Council of Ministers classified financial compensation into five levels based on vehicle fuel efficiency, as depicted in Tables 1 and 2. The implementation of this fee was carried out in two phases:

1. Phase I: This phase began in 2023 and applies to new light vehicles models in the year 2024.

2. Phase II: This phase began in 2024 and applies to all vehicle types.

**Table 1**. Proposed annual fees for light-duty vehicles made in 2015 and earlier and all heavy vehicles.

| Level | Engine size (L) | Fees (SAR) |
|---|---|---|
| First | < 1.9 | 0 |
| Second | 1.91–2.4 | 50 |
| Third | 3.2–2.41 | 85 |
| Four | 3.21–4.5 | 130 |
| Fifth | > 4.5 | 190 |

*Source: Markabati (SASO 2023).*

**Table 2.** Proposed annual fees for light-duty vehicles made in 2016 and afterwards.

| Level | Fuel consumption (km/L) | Fees (SAR) |
|---|---|---|
| First | > 16 | 0 |
| Second | 15.99–14 | 50 |
| Third | 13.99–12 | 85 |
| Four | 11.99–10 | 130 |
| Fifth | < 10 | 190 |

*Source: Markabati (SASO 2023).*

# 3. Model

A comprehensive analysis was conducted to examine the environmental implications associated with different fuel economy levels for light-duty vehicles in the KSA. This study focuses on gasoline-powered vehicle types. To facilitate the evaluation process, we utilized the alternative fuel life cycle environmental and economic transportation (AFLEET) tool (Argonne 2023). The AFLEET framework, developed by the U.S. Department of Energy (DOE), offers a comprehensive approach for assessing and comparing the energy and environmental performance of alternative fuel and advanced vehicle technologies. The AFLEET framework is a valuable resource for government agencies, businesses, and organizations seeking to make informed decisions about transitioning their fleets to alternative fuels. Its integrated approach considers both economic and environmental factors, aligning with the growing emphasis on sustainable and greener transportation solutions.

The vehicle categories in the AFLEET Tool are derived from the Motor Vehicle Emission Simulator (MOVES) developed by the Environmental Protection Agency (EPA 2022). They enable the estimation of emissions resulting from vehicle operation, including tailpipe emissions (EPA 2022).

## Inputs and Assumptions

We adjusted the input data within the AFLEET tool to relate to the appropriate factors that are prominent in Saudi Arabia. Our investigation focused primarily on crucial variables, as explained below:

- **Vehicle Type**: Passenger cars are four-wheel vehicles with two-axles that are primarily used for transporting passengers.
- **Car Sales (Historical and Future Projections)**: The average annual growth rate in car sales for the last five years, from 2018 to 2022, was 6.28%. If we exclude the COVID-19 year (2020), the growth rate for the last five years, from 2017 to 2022, becomes 4.60% (Saleh 2023). Therefore, in the business as usual (BAU) scenario, we assume an average 5% annual growth rate in car sales from 2023 until 2030. Figure 2 depicts the historical and future projections of car sales.

## Fuel Economy

The average fuel economy targets were obtained from the updated policy for the proposed Saudi Arabia corporate average fuel economy standard for new light-duty vehicles (2016-2020) (ICCT 2014), as presented in Table 3. Since the fuel economy targets for the years 2021 to 2030 were not published by the SASO, we proposed three scenarios, namely, 1%, 2%, and 3% yearly improvements in fuel economy until 2030. Figure 3 illustrates the historical fuel economy, KSA targets, and projected horizon years (ICCT 2014).

**Figure 2.** Car sales (historical and future projections): historical in light green, projected in dark green.

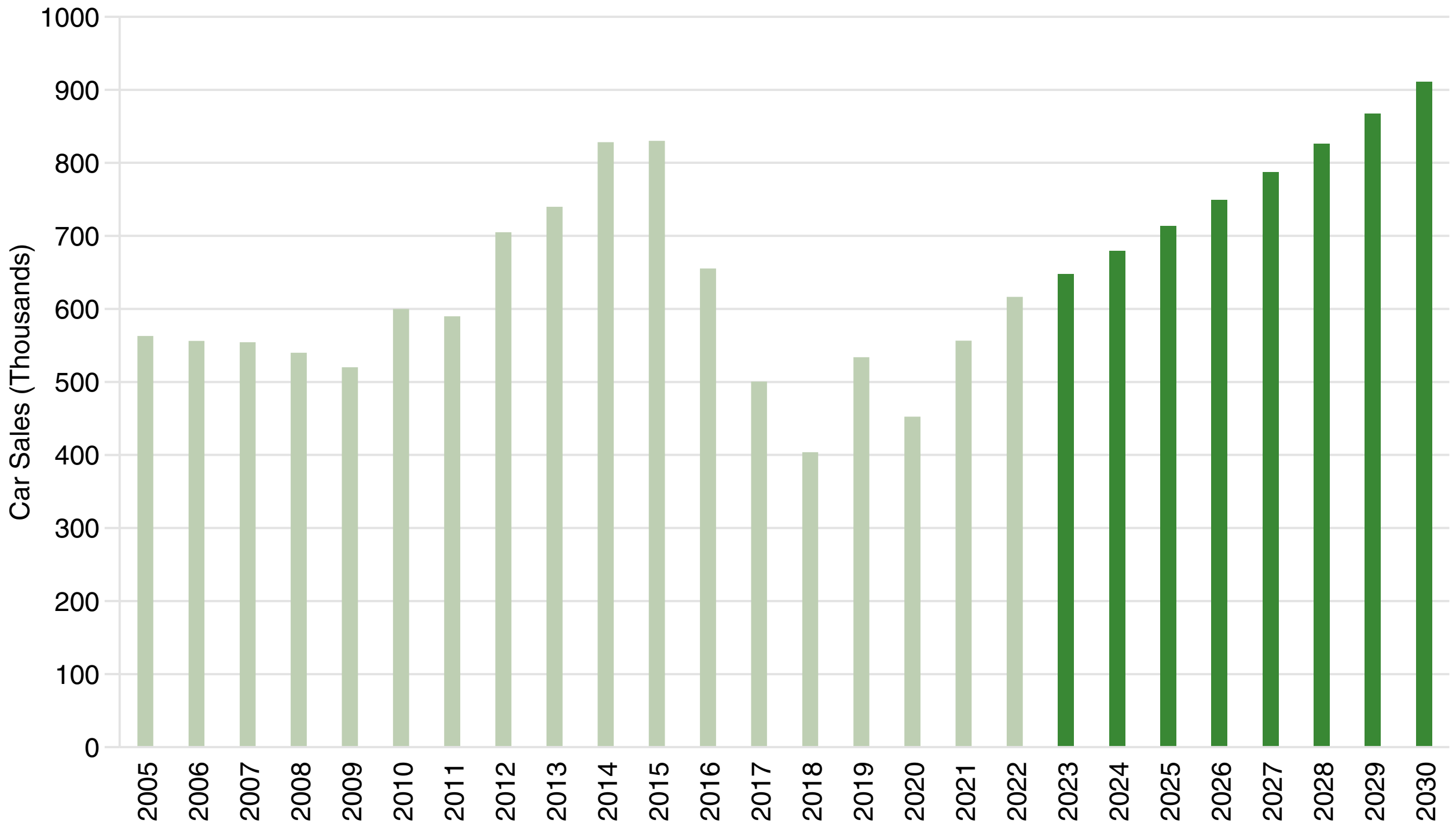

*Source: Statista (2023); Saleh (2023); Authors' Analysis.*

**Figure 3.** Fuel economy values for KSA: historical, targeted, and projected.

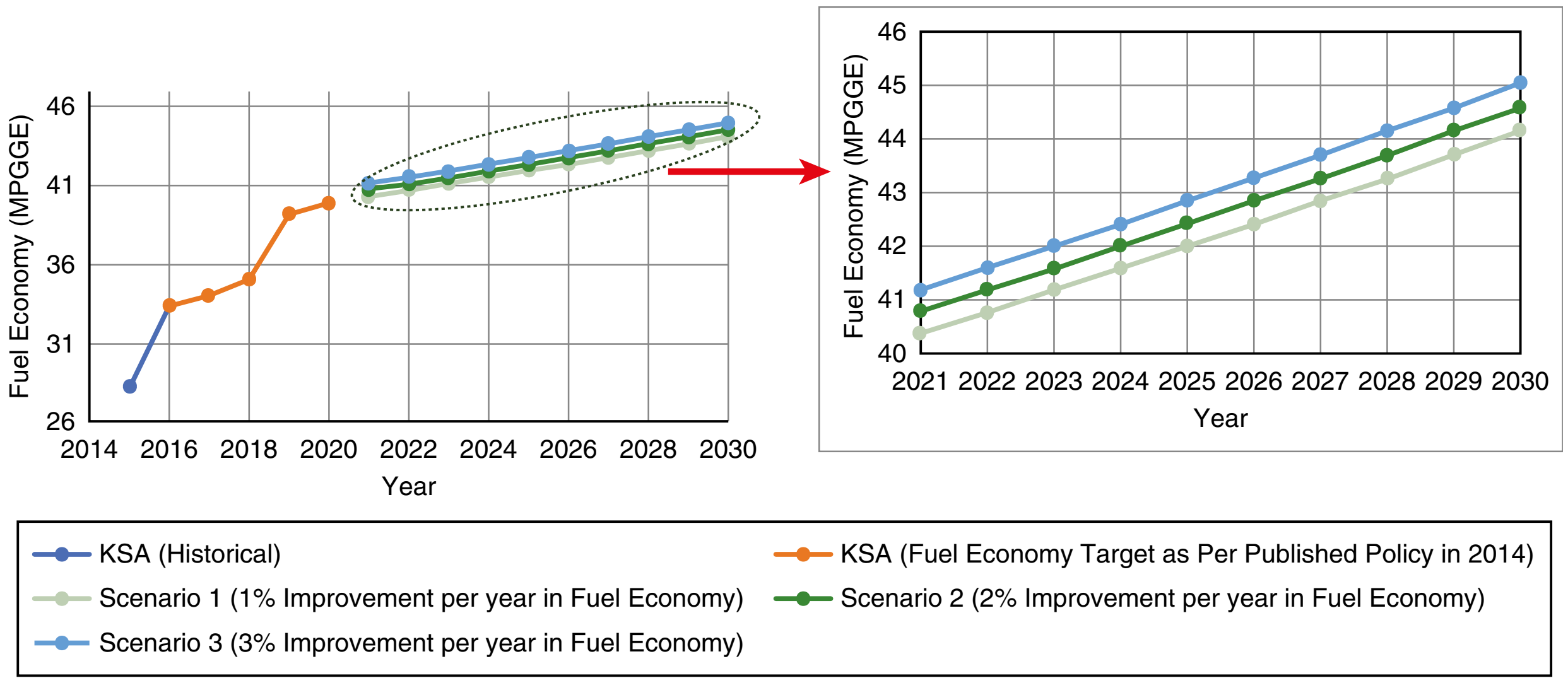

*Source: ICCT (2014); Updated Policy (2014); Authors' Interpretation.*

- **Average Annual Vehicle Mileage**: According to Sheldon (Sheldon 2019), the annual average mileage in the KSA is 16,000 miles. We accounted for the fact that this mileage may not remain constant due to the rebound effect from energy efficiency standards. This effect refers to changes in behavioral responses to fuel economy policies, which impact fuel savings and emission reductions. For instance, when vehicle fuel efficiency improves, driving costs per mile decrease, making driving less expensive; this situation can encourage more travel. This increase in driving due to reduced costs is known as the "direct rebound" effect.

  **Energy Consumption and Greenhouse Gas Emissions Calculations**: Well-to-Wheels Petroleum Use and GHGs & Vehicle Operation Air Pollutants Well-to-Wheels (WTW) analysis provides a comprehensive understanding of the entire lifecycle of petroleum use in the context of transportation. It encompasses the various stages involved, from the extraction of crude oil to the ultimate utilization of refined petroleum products in vehicles. For conventional internal combustion engine vehicles (ICEVs) running on gasoline, tank-to-wheel emissions typically account for the majority of WTW emissions, often approximately 80% (Zuccari 2019; Fueleconomy 2024). The cumulative GHG emissions in the study period (2016-2030) can be calculated using Equation (2).

$$GHG_j = \sum_{j=0}^{n} Car\ Sales_j \times VMT_j \times FE_j \times EF_j \tag{2}$$

where:

$GHG_j$ is the total GHG emissions in year *j*.

$Car\ Sales_j$ is the number of car sales in year *j*.

$VMT_j$ is the average annual vehicle mileage in year *j*.

$FE_j$ is the fuel economy target/consumption in year *j*.

$EF_j$ is the emission factor in year *j* (kg GHG per kg of fuel).

# 4. Empirical Results and Discussion

In this section, an in-depth exploration is conducted to assess the tangible impacts arising from the phase-in compliance mechanism entrenched within the fuel economy standard. The primary objective is to estimate the discernible effects pertaining to energy conservation and emissions reduction. This analysis is consistent with the predictions presented in Figure 3, which extend from 2016 to 2030. The specific horizon of this examination centers on the comparison between the average fuel economy values under the BAU scenario, measured at 28.2 miles per gallon gasoline equivalent (MPGGE), and the envisioned fuel economy targets, denoted as the Corporate Average Fuel Economy (CAFE) standard, presented in Figure 3. Importantly, this comparison is exclusively tailored for light-duty vehicles and confines its scope solely to the realm of annual car sales data.

The graphical representations summarized in Figures 4 and 5 illustrate the potential impact resulting from the adoption and enforcement of the proposed fuel economy standards. These illustrations clearly indicate a substantial decrease in both energy consumption and GHG emissions. Should the KSA adopt the phase-in compliance method in accordance with the policy update (ICCT 2014) of the CAFE standard for new light-duty vehicles, the actual fleet average (CAFE actual) would apply solely to the percentage of sales of its most efficient vehicles, as described in Table 3. According to the estimations, if the CAFE standards proposed in 2016 had been

**Table 3**. Phase in flexibility for compliance.

| EC no. | Compliance (%) |
|---|---|
| 1 | 80 |
| 2 | 90 |
| 3 | 100 |
| 4 | 100 |
| 5 | 100 |

*Source: ICCT (2014).*

followed from 2016 to 2020, both the cumulative energy consumption and GHG emissions for this period may have been 20% lower. Furthermore, under scenario 1, from 2021 to 2030, the cumulative energy consumption and GHG emissions may decrease by an additional 10%. Comparatively, in contrast to scenario 1, scenarios 2 and 3 show a marginal decrease of approximately 1.2% in both energy consumption and GHG emissions reduction.

**Figure 4.** Petroleum use of passenger car sales fleets (the left-axis represents annual values, and the right-axis represents cumulative values).

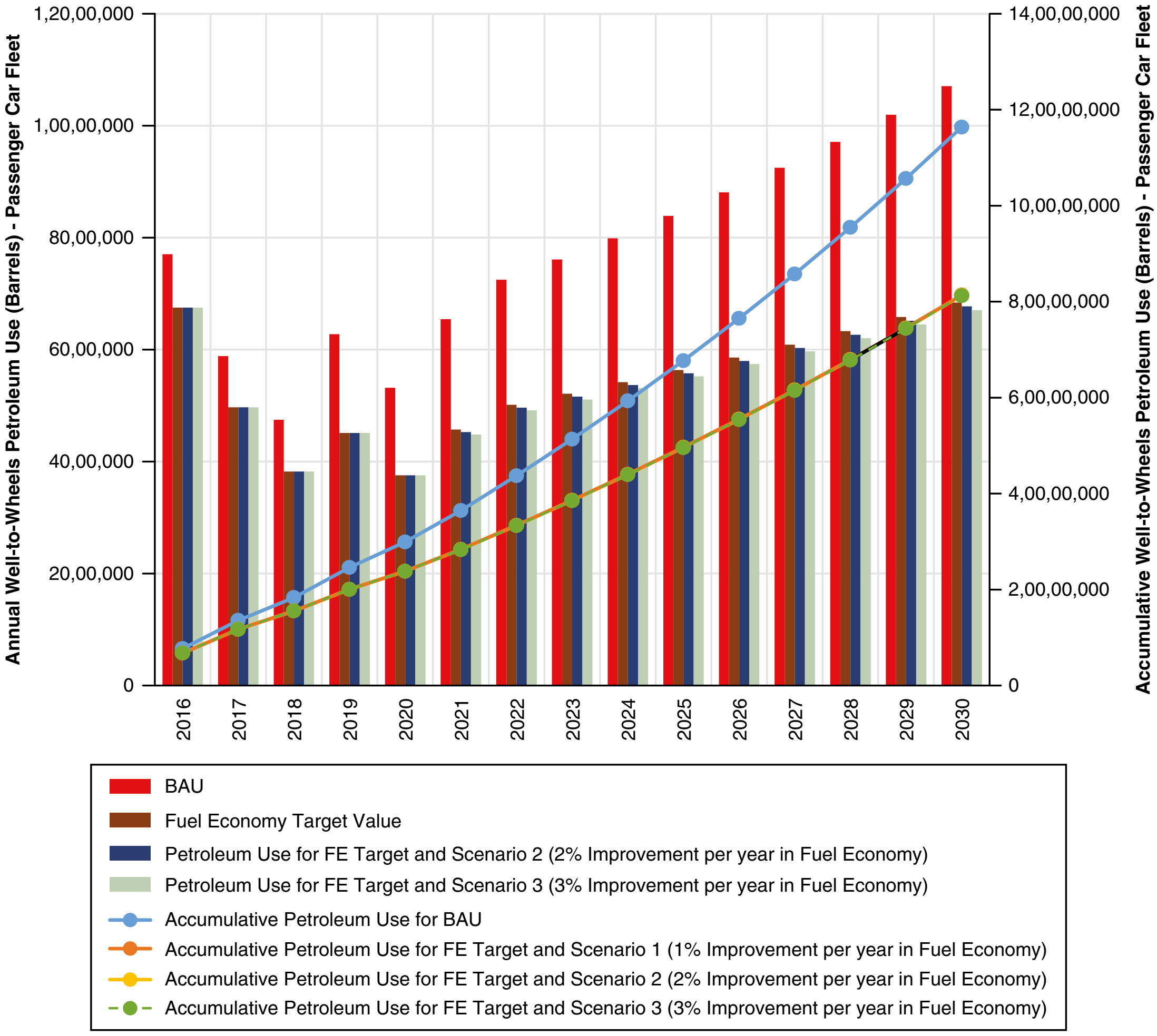

*Source: Authors' Analysis.*

**Figure 5.** GHG emissions[1] in short tons[2] for the passenger car sales fleet (the left-axis represents the annual values, and the right-axis represents the cumulative values).

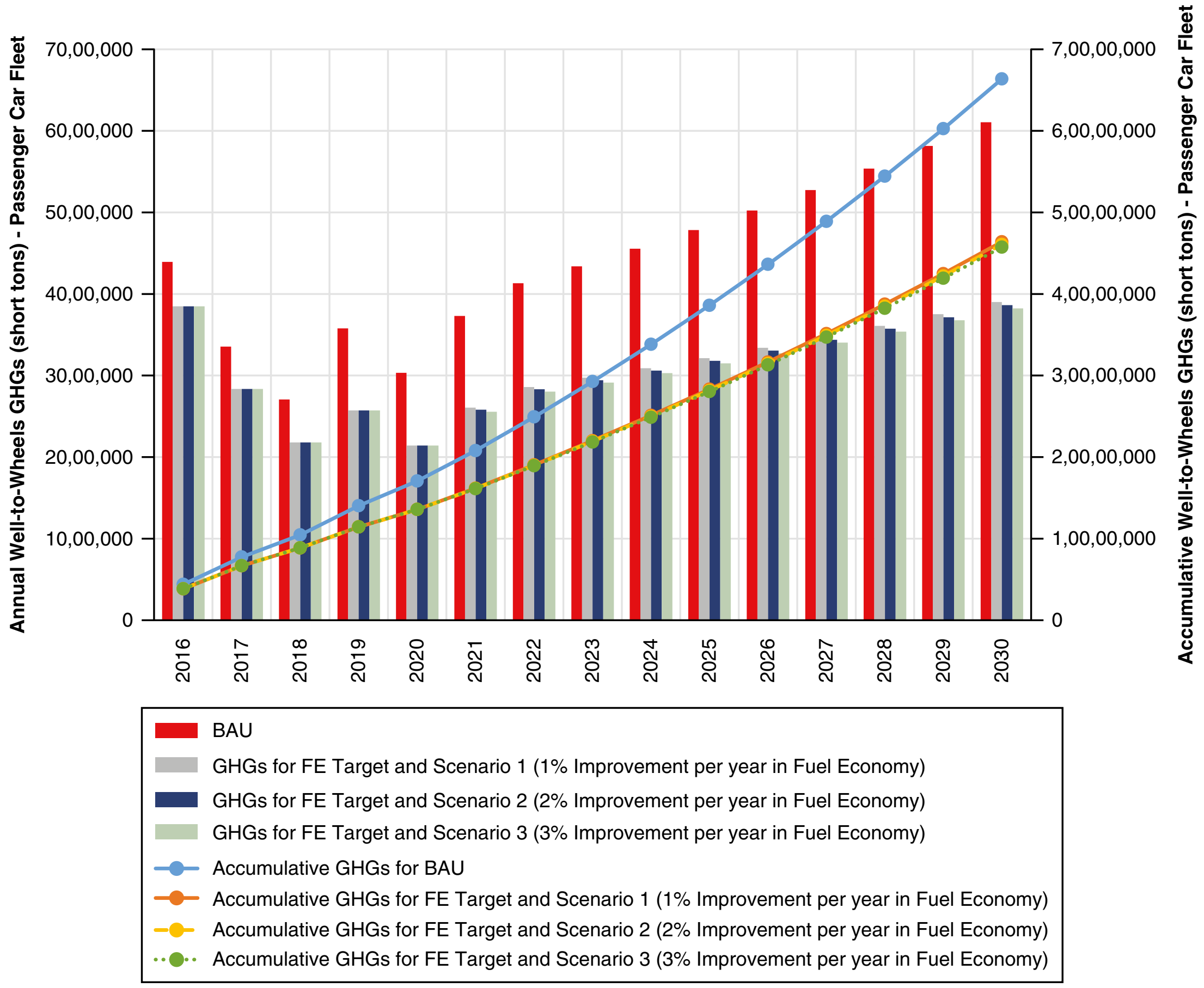

*Source: Authors' Analysis.*

# 5. Assessing the "Rebound Effect" from Fuel Economy Improvement

The rebound impact of energy efficiency policies, especially in relation to fuel economy standards, is characterized by behavioral and market reactions that might reduce the expected fuel savings and carbon reductions. Increasing a vehicle's fuel economy reduces the cost per mile, therefore making driving more economical. A decrease in cost can result in a greater frequency of driving, perhaps counteracting some of the environmental advantages obtained from enhanced efficiency. This phenomenon, in which a decrease in the cost per mile of driving results in an increase in the amount of driving, is commonly referred to as the "direct rebound effect" (Gillingham 2018).

In light of the limited data available in the KSA regarding the rebound effect of vehicle mile travel (VMT) due to improved fuel economy for light-duty vehicles, we follow the assumption made by (Gillingham 2018), who suggested that the rebound effect is approximately 10%.[3] Gillingham (2018) estimated that a 10% rebound effect in VMT due to fuel economy improvements serves as a valuable starting point for the KSA; however, several factors suggest that this figure may be relatively high for the local context in the KSA compared to that in the United States. First, because fuel prices are lower in the KSA than in the United States, the rebound effect could be minimized. With existing fuel subsidies already in place, consumers in Saudi Arabia may be less incentivized to alter their driving habits in response to improvements in fuel economy standards for vehicles, unlike their counterparts in the United States where fuel prices are typically higher (EIA 2024; ARAMCO, 2024). Moreover, data from The Royal Commission for Riyadh City (RCRC) indicate that more than 92% of daily trips in Saudi Arabia are made by private vehicles (UNDP 2022). This suggests that there may be limited potential for a significant shift from public transport to private vehicles due to improvements in vehicle fuel economy standards. As a result, the assumption of a 10% rebound effect may represent a worst-case scenario for the KSA. Figures 6 and 7 illustrate the increase in energy consumption and GHG emissions resulting from the rebound effect in scenario 3, which considers a 3% annual improvement in fuel efficiency from 2021 to 2023. An increase in fuel efficiency has a minor influence on energy consumption and GHG emissions, known as the rebound effect. In 2030, energy consumption and emissions are expected to increase by 4.5% compared to scenario 3, in which there is no rebound impact in vehicle mile travel. According to the statistics provided by Enerdata (2021), the transportation sector in Saudi Arabia had approximately 170 million metric tons of $CO_2$ emissions[4] in 2020. This figure is projected to increase to 288[5] million metric tons of $CO_2$ emissions by 2030 (Shannak et al. 2024). Adhering to the CAFE standards proposed in 2016 could have decreased the overall GHG emissions of the transportation sector by approximately 1.6% from 2016 to 2020. In addition, under scenario 3, GHG emissions may be reduced by an additional 4.16% by 2030.

**Figure 6.** Petroleum use of passenger car sales fleet with the rebound effect.

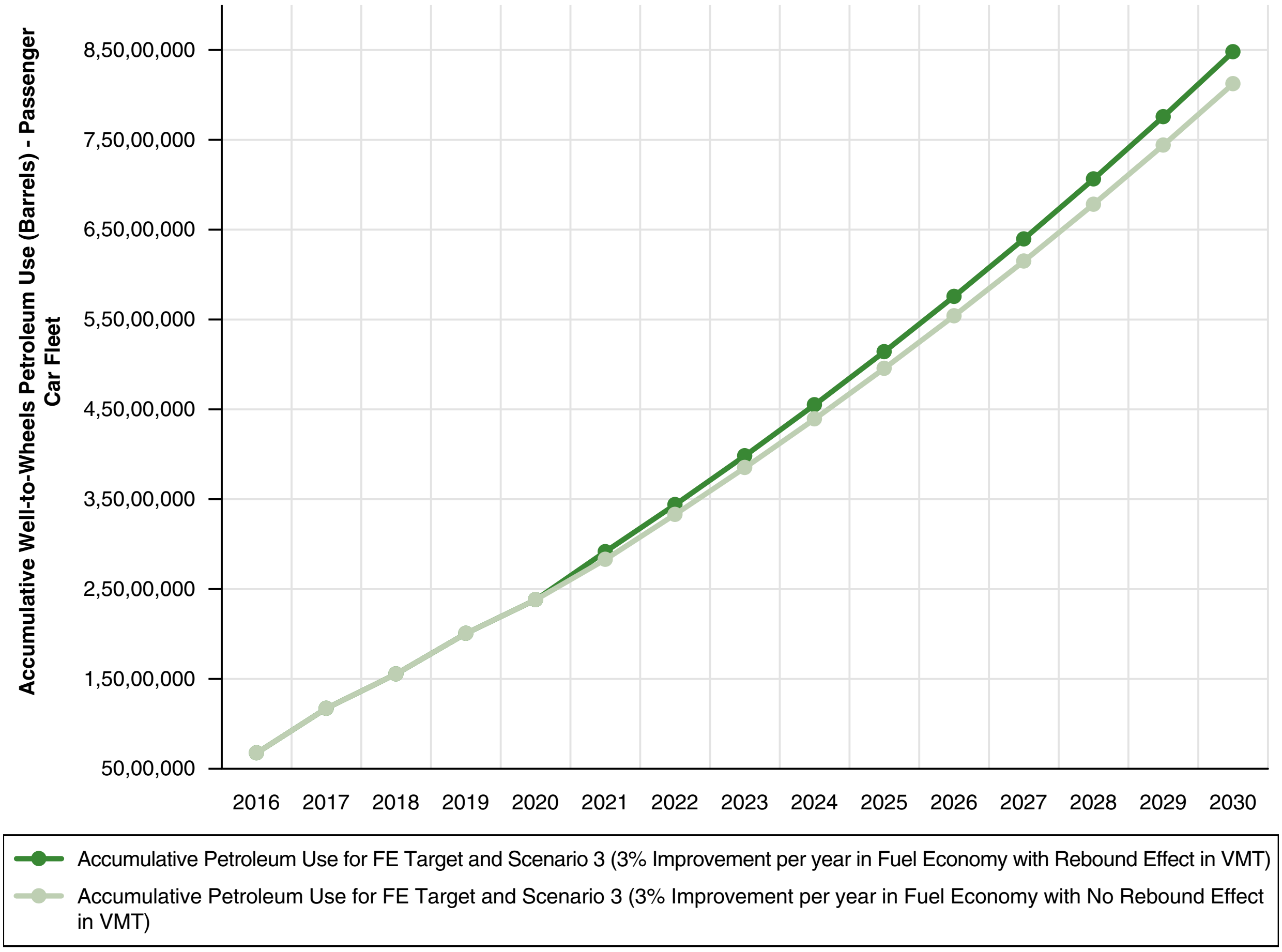

*Source: Authors' Analysis.*

**Figure 7.** GHG emissions[6] of the passenger car sales fleet with a rebound effect.

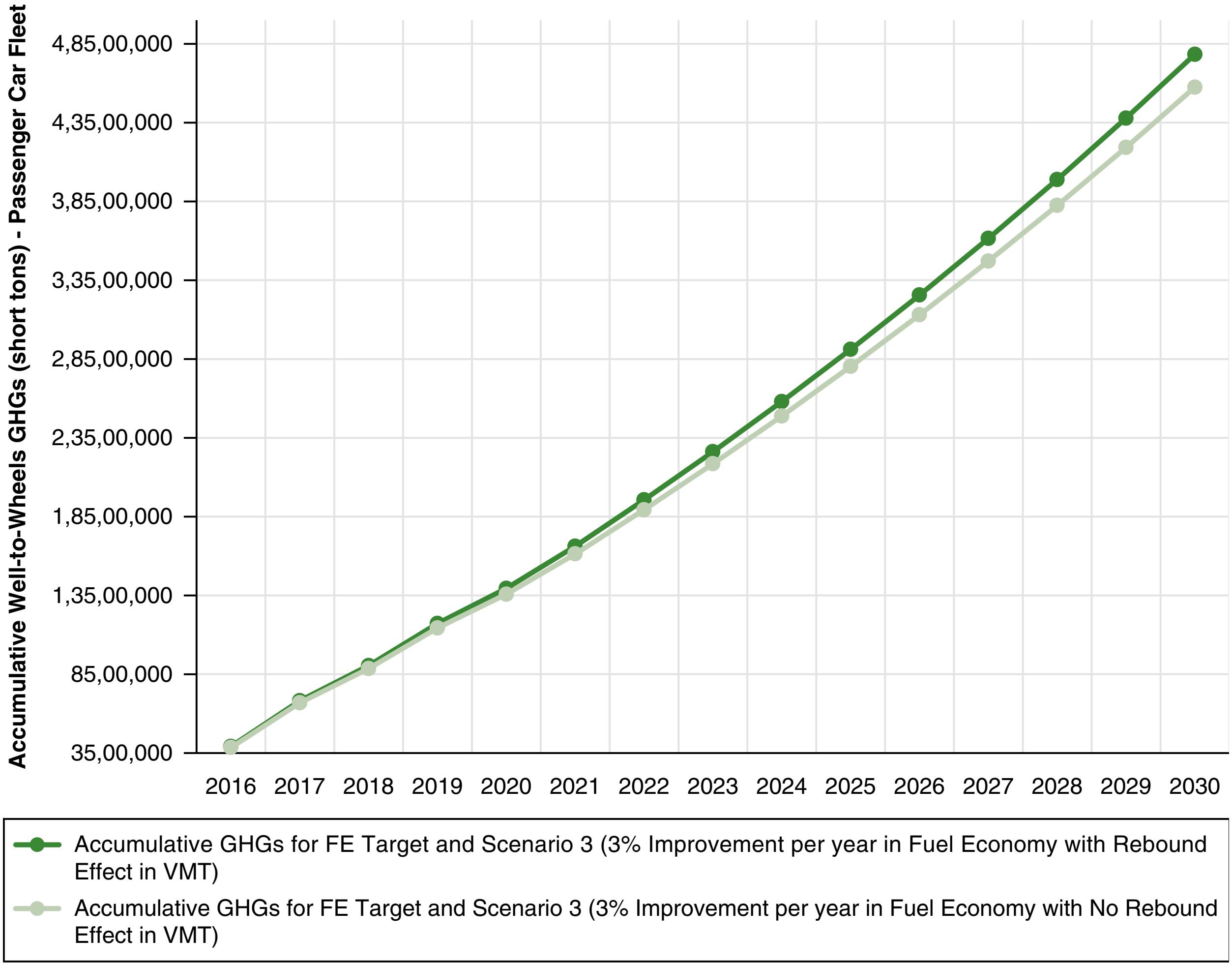

*Source: Authors' Analysis.*

# Conclusion and Policy Implications

This assessment of fuel economy targets within Saudi Arabia's transportation sector illustrates the significant potential for mitigating energy consumption and reducing emissions. The analysis, rooted in empirical analyses, elucidates the transformative impact of adhering to proposed fuel economy standards. Should Saudi Arabia implement the phase-in compliance method aligned with the policy update (ICCT 2014) regarding the CAFE standard for new light-duty vehicles, the actual fleet average (CAFE actual) would specifically pertain to the proportion of sales associated with its most efficient vehicles. Estimations show that if the CAFE standards proposed in 2016 had been followed from 2016 to 2020, both the cumulative energy consumption and GHG emissions for this period would have been 20% lower. Furthermore, from 2021 to 2030, the cumulative energy consumption and GHG emissions could decrease by an additional 10% under scenario 1. Comparatively, in contrast to scenario 1, scenario 3 shows a marginal decrease of approximately 1.2% in both energy consumption and GHG emissions reduction. In terms of the total GHG emissions from the transport sector, from 2016 to 2020, adherence to the CAFE standards proposed in 2016 could have decreased emissions by approximately 1.6%. Moreover, scenario 3, which assumes 3% efficiency per annum from 2021 to 2030, could enable a further decrease of 4.16% in GHG emissions by 2030. The adoption of more stringent fuel economy standards, as evidenced in this study, represents a pivotal pathway toward achieving net zero emissions in the transportation sector. By steadily increasing the efficiency of vehicles, this approach drives a gradual but consistent decline in emissions. When coupled with the simultaneous integration of electric and alternative fuel vehicles into the market, the goal of net zero emissions becomes increasingly feasible.

Based on the findings and the practices gathered from the implementation of fuel economy standards, several recommendations and policies aimed at aligning light-duty vehicle efficiency with GHG emissions to achieve net zero emissions in the transportation sector merit consideration:

1. **Scaling Up Fuel Economy Standard Targets**

    - Future targets for fuel economy standards that promote the adoption of more efficient technologies in conventional internal combustion engine vehicles should be established.
    - Vehicle sales share targets for alternative fuels (hybrid, electrical, plug-in hybrid) should be set to encourage the market penetration of electric and alternative fuel vehicles.

2. **Compliance Assessment, Monitoring, and Enforcement**

    - Robust monitoring and enforcement mechanisms should be implemented to ensure that manufacturers adhere to set standards and targets. This could involve regular audits, inspections, and penalties for non-compliance.
    - Ongoing evaluation and adjustments to standards and regulations based on technological advancements, market trends, and environmental goals should be facilitated.

Implementing these recommendations and policies will contribute significantly to achieving net zero emissions in the transportation sector by steadily increasing vehicle efficiency and accelerating the adoption of low-emission and zero-emission vehicles.

Although the current study concentrates solely on gasoline vehicles, our upcoming research intends to undertake a thorough evaluation of the impact of vehicle penetration rates pertaining to different fuel types (namely, gasoline, hybrid, plug-in hybrid, and electric vehicles), electricity generation sources, and the introduction of public transportation. This assessment will specifically examine the effects of energy conservation and emissions within the context of the KSA.

# Endnotes

[1] AFLEET reports GHGs as $CO_2$.

[2] 1 short tons = 0.90718474 metric tons. equivalent.

[3] In the context of fuel prices, if there's a 10% rebound effect, it means that for every 1% decrease in fuel prices, there is a 0.1% increase in vehicle miles traveled (VMT).

[4] The $CO_2$ emission is typically the largest component of GHG emissions accounting for 80% of the total GHGs in Saudi Arabia, based on 2021 data (Climate Watch 2023). We used the 2020 $CO_2$ emissions value from Enerdata (2021), which is 118.3 Mt. Then, to get the total GHG emissions this number is increased by 20%, according to the Climate Watch (2023) information. To end up with a well-to-well number, the resulting number is also increased by 20% based on Zuccari (2019) and Fueleconomy (2024).

[5] The number is calculated as in Footnote 4, using the projected number by Shannak et al. (2024).

[6] AFLEET model reports GHGs as $CO_2$ equivalent.

# About the Project

Mapping the Net-Zero Pathways to Sustainable Transport in Saudi Arabia. This project aims to create an evidence-based, comprehensive understanding of the different projects, policies and regulations related to international connectivity, regional transport and urban mobility in Saudi Arabia; their contribution to energy demand and lowering of GHGs and emissions; and the establishment of a net-zero pathway for the KSA.

مركـــز الملـــك عبـــدالله للدراســـات والبحوث البتروليـــة
King Abdullah Petroleum Studies and Research Center

www.kapsarc.org